# Comparing lifetime and annual fitness measures reveals differences in selection outcomes

F. Stephen Dobson[1,4], Claire Saraux[1*], David W. Coltman[2],

Shirley Raveh[3] & Vincent A. Viblanc[1*]

[1]Université de Strasbourg, CNRS, IPHC UMR 7178, F-67000 Strasbourg, France

[2]Department of Biological Sciences, University of Alberta, Edmonton, AB, Canada

[3]School of Biodiversity, School of Biodiversity, One Health and Veterinary Medicine,

University of Glasgow, Glasgow, UK

[4]Department of Biological Sciences, Auburn University, Auburn, AL, USA

§ Correspondence: fsdobson@msn.com

* These authors contributed equally to the study

# ABSTRACT

Selection analyses of long-term field data frequently use annual comparisons from long-lived species with overlapping generations to measure fitness differences with respect to phenotypic characteristics, such as annual phenological timing. An alternative approach applies lifetime estimates of fitness that encompass several annual events. We studied selection on emergence date from hibernation in male and female Columbian ground squirrels, *Urocitellus columbianus* (Ord 1915). From 32 years of records, we estimated lifetime fitness using either lifetime reproductive success (LRS) or matrix methods, and estimated annual fitness from individual yearly survival and reproduction. We also modified estimates to statistically control for changes in mean population fitness over the lives of individuals. We regressed lifetime fitness metrics on dates of emergence from hibernation, to quantify the strength of selection on emergence date (a heritable trait). All fitness metrics were highly correlated, but differences became apparent when estimating selection coefficients for emergence dates. The annual fitness metric and LRS produced lower effect sizes of selection coefficients than matrix-based lifetime fitness and a lifespan approach based on average annual fitness. Further, only these last two metrics revealed stabilizing selection. These results suggest that the choice of a fitness metric may influence our conclusions about natural selection.

# INTRODUCTION

Fitness differences among trait-based groups within populations are an important component of studies of natural selection in the wild (e.g., Endler 1986; Charmantier et al. 2014). Ideally, one examines changes in traits under selection in natural populations from one generation to the next. This measure of trait change would include influences of both survival and reproduction at different times in the lives of individuals, and would not necessarily reflect the specific life events that led to changes in trait frequencies. The importance of specific life events is that they can be directly linked to the agents of selection that cause changes in trait frequencies via changes in reproduction and survival (Wade and Kalisz 1990). These selective agents are specific factors in the social or ecological environment that influence reproduction and survival, including abiotic factors such as climate, and biotic factors such as predators, parasites, important environmental resources, or the density of conspecific competitors. Naturally, the interplay of ecological interactions and evolutionary outcomes is an important topic for both ecologists and evolutionary biologists (Hutchinson 1965).

Several difficulties become evident when trying to execute a research program that measures evolution by natural selection. The first is choosing a trait that is heritable. Another is that there may be genetic correlations among traits, termed "the G matrix" for the variance-covariance relationships among heritable traits (Hadfield 2008). The evolution of a particular trait form may be misunderstood when the G matrix is unknown, as is the case of most studies of natural selection in the wild (Arnold et al. 2008; Engen and Sæther 2021). Yet another problem is that if the trait(s) of interest is measured at expression in adults, any estimation of selection on the trait in adulthood may suffer from earlier selection before the trait is expressed (termed the "invisible fraction" Hadfield 2008; Mittell and Morrissey 2024). Finally, measures of selection on alternative forms of a trait may depend on whether fitness is measured on an annual or lifetime basis. For example, in a study of sociality and fitness in house sparrows (*Passer domesticus* Linnaeus, 1758), Dunning et al. (2023) found that birds with more opposite-sex associations had higher annual fitness, but not lifetime fitness.

Previous studies have examined fitness differences that were measured on an annual or even shorter-term basis, though data may have been obtained from several years (e.g., Lande and Arnold 1983; Kruuk et al. 2002; Dobson et al. 2017). Dobson et al. (2020) suggested that annual measures should be most appropriate for experiments and short-term events, such as annual phenology. However, for long-lived species, such events can be repeated over the lifetime and the realized fitness of these traits may take a complete lifetime to be observed. An individual with a particular trait expression might be favored by selection in one year, but dis-favored the next (Sæther and Engen 2015). As well, plastic trait expression may change with environmental conditions from year-to-year in unexpected ways. This seems especially likely if there are carry-over effects or life-history tradeoffs from one year to the next, such as an allocation of resources like body fats or proteins to reproduction in one year that leaves fewer residual resources for the following year. While alternative approaches may examine somewhat different questions (viz., a specific year versus several years of a lifetime), it is not clear for long-term datasets whether an annual measure taken over several years or lifetime records better reflect the influence of events that vary both within years and among years for specific individuals.

Our goal was to apply lifetime fitness measures from a 32-year study of Columbian ground squirrels (*Urocitellus columbianus* (Ord, 1815)) to a specific trait of males and females, the date at which they emerge above ground from their annual period of hibernation, and compare this measure to selection gradients on emergence date obtained from annual measures of fitness (as in Viblanc et al. 2022). First, we estimated lifetime fitness using either lifetime reproductive success (LRS) or matrix methods introduced by McGraw and Caswell (1996), and estimated annual fitness from individual yearly survival and reproduction following Qvarnström et al. (2006). We then modified these estimates to statistically control for changes in mean population fitness over the lives of individuals. In a growing population, estimated fitness for a trait form will be relatively lower than in a declining population when compared to other trait forms, so we calculated estimates of relative fitness (see Viblanc et al. 2010; Dobson et al. 2020; Methods). We compared associations among lifetime fitness metrics when controlling statistically or not for changes in the mean fitness in the population. Second, we compared selection gradients using lifetime fitness to selection gradients obtained from annual fitness (updated from Viblanc et al. 2022), to discern any differences between lifetime and short-term measures when used in analyses of natural selection.

For our measure of annual or lifetime fecundity, we focused on number of offspring that survived until yearling age, as this appeared to be a reasonable first approximation to measuring fitness from offspring that survive to pass on traits (Viblanc et al. 2022). We expected that since estimates of annual and lifetime fitness were only moderately associated (average $r = 0.40$; Dobson et al. 2020), the different calculations for selection gradients with respect to the single trait of emergence date might also differ. We did not examine multigenerational estimates, since these are seldom available in field studies (but see Reid et al. 2019; Alif et al. 2022; Van de Walle et al. 2022). We regressed lifetime fitness estimates or annual fitness estimates on linear and quadratic terms for emergence dates for males and females, to test for stabilizing or disruptive selection (Lande and Arnold 1983; Arnold and Wade 1984).

Columbian ground squirrels are medium-sized rodents that have an extremely long hibernation period (about 265 days) and short active season in which to reproduce and store resources for subsequent hibernation. They are also fairly long-lived, with a median age of females at annual spring emergence from hibernation of 3 (range = 1-14 years old, n = 1277) and a median age of males of 2 (range = 1-10 years old, n = 787), and thus have overlapping generations. Emergence dates from hibernation exhibit a strong genetic correlation with estrous date ($r_G$ = 0.98 ± 0.01 standard error; Lane et al. 2011), the latter an important reproductive characteristic (Dobson et al. 1999). We could not separate direct and indirect selective influences on emergence date, but chose to examine fitness differences with emergence date due to larger sample sizes. Adult males emerge from hibernation about 7-10 days before adult females (Murie and Harris 1982; Dobson et al. 1992; Tamian et al. 2022; Thompson et al. 2023). Emergence date is also an appropriate trait to examine because it exhibits significant heritability in both males and females (respectively, $h^2$ = 0.34 and 0.22; Lane et al. 2011). Finally, emergence date is also a highly plastic trait that exhibits changes from year to year (Dobson 1988; Lane 2012; Dobson et al. 2016, 2024; Tamian et al. 2022; Thompson et al. 2023).

## Materials and Methods

*Study species and long-term population monitoring*

From 1992 to 2023, we studied Columbian ground squirrels in a single "pocket meadow" in the Sheep River Provincial Park, Alberta, Canada (50° 38′ 10.73″ N; 114° 39′ 56.52″ W; 1524 m; 2.3 ha). Details of animal handling and monitoring are presented elsewhere (Viblanc et al., 2022). Of importance here is that individual ground squirrels were marked for permanent recognition with numbered metal ear tags (National Band and Tag Company, Newport, KY) when weaned (or at first capture for immigrants) and dyed annually with a unique black mark on the pelage with human hair dye (Clairol® Hydrience, Clairol Inc., New York, USA). Monitoring included trapping the entire population at spring emergence, using peanut butter bait and live-traps (chipmunk size, Tomahawk Live Trap, Hazelhurst, WI, USA). We followed ground squirrels daily throughout their breeding season. Note that due to the COVID-19 pandemic, data on emergence dates could not be collected in 2020, as we arrived too late in the field.

Females mate, frequently with multiple males, usually 3-5 days after emergence from hibernation (Murie and Harris 1982; Raveh et al. 2010). About 24 days later, mothers give birth in a single-entrance "nest burrow" (Murie et al. 1998). For about 27-28 days after birth, mothers lactate for their young (Murie and Harris 1982), and then the litter emerges above ground from the natal nest burrow. We again caught and examined the mother at this time, and caught and ear-tagged her entire litter. All ground squirrels subsequently fatten (the young also continue to develop) for the long hibernation period, and male and female young remain in the natal colony for their first hibernation. Yearling males generally disperse before young of the year are weaned, though some remain residents on their natal meadow (Neuhaus 2006). Virtually all females remain as residents on their natal meadow (Wiggett and Boag 1992).

*Paternity analyses*

A small sample of ear tissue was taken from the outer pinnae of immigrant males, the first time they were seen and captured as immigrants in the population. Tissue samples were taken from a toe bud of all young at birth from 2005 to 2016 (described in Viblanc et al. 2022) or from the ear pinnae as described for adults in other years. Tissue was also sampled from the ears of a few adults that had not been previously sampled. Paternity was estimated using the tissue samples from adults and young in 2001-2021, analyzed for 13 microsatellites, and assigned using CERVUS 3.0 (Marshall et al. 1998; Kalinowski et al. 2007), as described in detail by Raveh et al. (2010). Dams were known for all offspring from association at birth and from cohabitation in the natal burrow. Sires were assigned with 95-99% confidence and with 2 or fewer alleles that did not match either or both parents (such maternal mismatches revealed infrequent allelic mutations). Because of an adult sex bias towards females in the population and because paternity analyses were not available in all years, the dataset on male fitness is smaller than for females.

*Lifetime fitness*

We calculated individual lifetime fitness following McGraw and Caswell (1996) for all individuals that reached maturity (observed mating at least once for females and observed with pigmented scrotal testicles for males). Briefly, this approach mirrors the Leslie's fecundity-survival projection matrix (Leslie 1945) approach to calculate population growth rate, and applies it at the scale of an individual's survival and reproduction. For a pre-breeding population census (after Oli and Zinner 2001), the population size $N$ at $t+1$ for age groups 0, 1 up to $\delta$ is given by:

$$\begin{bmatrix} N_{1,t+1} \\ N_{2,t+1} \\ \vdots \\ N_{\delta-1,t+1} \\ N_{\delta,t+1} \end{bmatrix} = \begin{bmatrix} F_1 & F_2 & \ldots & F_{\delta-1} & F_\delta \\ S_1 & 0 & \ldots & 0 & 0 \\ \ldots & \ldots & \ldots & \ldots & \ldots \\ 0 & 0 & S_{\delta-2} & 0 & 0 \\ 0 & 0 & \ldots & S_{\delta-1} & S_\delta \end{bmatrix} \times \begin{bmatrix} N_{1,t} \\ N_{2,t} \\ \vdots \\ N_{\delta-1,t} \\ N_{\delta,t} \end{bmatrix}$$

where $S_{1,\ldots,\delta}$ are the survival rates for age groups $1, \ldots, \delta - 1$ (*survival of* $N_\delta$ *is* 0), and $F_{1,\ldots,\delta}$ is the average fertility defined as the number of offspring produced that survived until emergence after their first hibernation divided by 2, since the analyses are conducted separately on the male and female part of the population (Dobson and Oli 2001).

The dominant eigenvalue of the Leslie fecundity-survival matrix provides us with the finite asymptotic growth rate of the population, lambda ($\lambda$). When $\lambda$ is greater than 1, the population grows and when it is lower than 1, the population decreases. Adapting the Leslie matrix to the scale of the individual, one can use the dominant eigenvalue as a measure of individual lifetime fitness ($\lambda_{ind}$, *i.e.* the growth of individual genotypes over life). The Leslie matrix then becomes:

$$\begin{bmatrix} F_1 & F_2 & \ldots & F_{\delta-1} & F_\delta \\ 1 & 0 & \ldots & 0 & 0 \\ 0 & 1 & \ldots & 0 & 0 \\ 0 & 0 & \ddots & 0 & 0 \\ 0 & 0 & \ldots & 1 & 0 \end{bmatrix}$$

Here also, the fecundity of a given individual across its different years of life is calculated as the number of offspring produced, and divided by 2 since only half of the genes passed on to the offspring are from maternal or paternal contribution. In this case however, survival is binary: 1 in years when the individual survived, and 0 when the individual died and $\delta$ is the maximum age of the given individual. To match the population pre-breeding census, we calculated individual fecundity by counting the number of offspring produced in a given year that survived to yearling age, also in accordance with Viblanc et al. (2022).

Similarly to population ecology, where a $\lambda$ of 1.0 indicates a stable population (Caswell 2001), a $\lambda_{ind}$ of 1.0 simply means that the individual replaces itself. Yet, an individual with a $\lambda_{ind}$ of 1.0 would be below average in an increasing population, but above average in a decreasing population. Because fitness is relative to what others are doing, a fair comparison of $\lambda_{ind}$ between different individuals needs to account for changes in demography over individual lifetimes

(Viblanc et al. 2010; Dobson et al. 2020; Rubach et al. 2020). We thus adjusted our individual fitness estimates so that a $\lambda_{ind_{adj}}$ of 1.0 would represent a stable genotype in the population (its proportion staying the same). To do so, for each individual a population Leslie matrix was estimated over its lifetime (from age 1 to death, as we were doing pre-breeding census), where population demography was pooled over age groups during the total lifespan of the individual. Then population growth was estimated as the dominant eigenvalue of this matrix. $\lambda_{ind_{adj}}$ equaled the ratio of $\lambda_{ind}$ over the population growth. This required estimation of age-dependent fecundity and survival. Due to our high recapture effort, we assumed a recapture probability of 1 for females and adult males. However, because some males disperse between age 1 and 2, the estimation of the age 1 survival was made possible only by assuming that emigration = immigration. Male age 1 yearly survival was thus estimated as the ratio between the number of age 2 males present in year y+1 divided by the number of age 1 males present in year y.

We estimated LRS by simply summing lifetime numbers of offspring that survived their first winter, produced by both males and females that reached maturity and had complete lifespans on our study meadow. We statistically controlled for changes in LRS-estimated mean population fitness over individual lifetimes by dividing individual LRS by the mean LRS of the individual's cohort. Note that our estimates could be done only for individuals born in cohorts for which all individuals perished during the study period. This decreased the sample size, especially in females, which included all cohorts for females born in 1992 to 2011 and 2013, and for males with complete lifetime records that were sampled for paternity from 2005 to 2017.

*Annual fitness*

To compare lifetime and annual measures, we updated the data published in Viblanc et al. (2022) through the 2023 field season. Annual fitness ($\omega_{an}$) was calculated following (Qvarnström et al. 2006) as individual survival (1/0) plus half of the annual breeding success (i.e. number of offspring that survived to yearling age). Annual breeding success was halved since only half of an individual's genetic contribution is passed on to offspring. Annual fitness was only estimated for mature individuals (i.e. females observed mating and males with scrotal testes and pigmented scrotum). The index represents the number of gene copies in the next year. We accounted for changes in demography ($\omega_{an_{adj}}$) by dividing individual annual values by the mean individual fitness of the population for a given year (Lande and Arnold 1983). Annual values (either raw or adjusted) were then averaged over the individual's lifetime ($\bar{\omega}_{life}$ and $\bar{\omega}_{life_{adj}}$) to compare with other lifetime fitness metrics.

*Selection analyses*

In our test for directional, stabilizing, or disruptive selection on emergence date from lifetime fitness, we centered the date by subtracting the mean from each value in each year. We confirmed that centered emergence dates were significantly repeatable (ICC = the intraclass correlation coefficient), albeit at a low effect size for females (ICC = 0.15, CI$_{95}$ = [0.09 – 0.22], n = 1092 observations, N = 257 females, P ≤ 0.001) and males (ICC = 0.17, CI$_{95}$ = [0.07 – 0.27], n = 417 observations, N = 113 males, P < 0.001), and calculated the mean-centered emergence date for each individual over its breeding years. This could not be estimated for 4 individuals (3 females and 1 male) that only bred in 2020, when no data on emergence dates were available due to late arrival in the field associated with the COVID-19 pandemic.

We then tested the selection gradient acting on emergence date by separately regressing our fitness metrics on: (1) mean-centered emergence date or (2) mean-centered emergence date and (mean-centered emergence date)$^2$, in males and females separately, using linear models. The general form of these models was:

$$\theta = \alpha + \beta z + \varepsilon \quad (1)$$

$$\theta = \alpha + \beta z + 1/2\gamma z^2 + \varepsilon \quad (2)$$

where $\theta$ is the measure of fitness ($\lambda_{ind_{adj}}$, LRS$_{adj}$, $\bar{\omega}_{life_{adj}}$) under consideration, $\alpha$ is the intercept, $z$ is the phenotypic trait of interest (viz., centered emergence date), and $\varepsilon$ is an error term. Directional selection was indicated by significant linear coefficients ($\beta$), the sign of the coefficients indicating the direction of selection. Stabilizing or disruptive selection occurs when the quadratic ($\gamma$) coefficient is significant (Lande and Arnold 1983; Arnold and Wade 1984; McGraw and Caswell 1996). Here, we report selection coefficients as recommended by Lande and Arnold (1983), which do not require doubling of the quadratic term to be interpreted (Stinchcombe et al. 2008).

We ran the analyses separately for males and females because the variance of fitness metrics was far greater in males than females (Viblanc et al. 2022; see also Jones et al. 2012). All fitness estimates were first standardised to a mean of zero and unit standard deviation before being processed in the model, so that effect sizes could be compared. For a matter of comparison, we also updated and standardized our estimates of selection gradients obtained with annual measures from Viblanc et al., 2022; using a mixed model with both ID and AGE included as random factors. We note that $\lambda$ and $\omega$ are measured as a growth rate during one year, while LRS is, on average, measured over a generation (one generation for females is 2.4 - 3.8 years, Zammuto 1987). Finally, the age at which males and females mature at first breeding is variable

(for females, Rubach et al. 2020). Thus, our analyses include only individuals that reached maturity (see above).

*Statistics*

All analyses were done in R v. 4.0.2 (R Core Team 2020). The repeatability of annual fitness or emergence date was estimated using the 'rptR' package in R (Stoffel et al. 2017), using a normal distribution and the calculation of an intraclass correlation coefficient (ICC) = $\frac{V_I}{V_P} = \frac{V_I}{V_I + V_R}$, where $V_I$ is the among-individual variance, $V_R$ is the within-individual (or residual) variance and $V_P$ is the total variance in fitness. Repeatability analyses were conducted using data only from individuals observed at least twice. Correlation between annual fitness, lifetime fitness and LRS were run using Spearman rank correlation tests, as not all distributions were Gaussian. All means were reported $\pm 1$SE. Because LRS could not be adjusted for changes in mean LRS-based population fitness for all individuals (only for those of cohorts in which all individuals perished within our total study period), samples sizes vary between models and are indicated as n (the number of breeding events), $N_1$ (the number of individuals), and $N_2$ (the number of years).

## RESULTS

**Male and female fitness**

Females and males that reached maturity lived a similar amount of time on average (4.3 ± 0.2 years, range: [1-14], N = 209; 4.1 ± 0.3 years [2-10], N = 56; for females and males respectively). However, males displayed higher and more variable fitness metrics. Male lifetime reproductive success (LRS) averaged 5.6 ± 1.1 pups that survived their first winter [0-36], while female LRS averaged only 2.8 ± 0.2 [0-15]. Lifetime fitness based on the individual matrix approach, $\lambda_{ind}$, was 0.90 ± 0.09 [0.00 – 2.18] for males *vs.* 0.77 ± 0.04 [0.00 – 2.00] for females. Finally, average annual fitness over an individual's lifetime, $\overline{\omega}_{life}$, was 1.24 ± 0.15 [0.00 – 4.88], *vs.* 0.85 ± 0.04 [0.00 – 1.93], for males and females respectively.

All three lifetime fitness metrics (LRS, $\lambda_{ind}$, $\overline{\omega}_{life}$) calculated from raw data, *i.e.,* not controlling for mean population fitness over individual lifetimes, were highly correlated (r ≥ 0.96; Table 1A; Fig. 1). While these correlations slightly waned when we controlled for changes in mean population fitness, they still remained high (r ≥ 0.78; Table 1B; Fig. 2). Correlations with LRS, whether adjusted for changes in estimated mean population fitness or not, appeared to have a curvilinear association with matrix-based or averaged annual fitness measures (Figs. 1 and 2).

**Shape of selection on emergence date according to fitness metrics**

We first tested for stabilizing or disruptive selection using the different fitness measures (Table 2). We found no significant selection (neither directional, disruptive nor stabilizing) on emergence dates when considering adjusted LRS, either in males or in females (Fig. 3 C&D, Table 2). However, the other lifetime fitness metrics yielded significant selection on emergence dates. Adult females experienced stabilizing selection on emergence from hibernation, with a peak about 3 days before the mean emergence date (thus, directional and stabilizing) when using lifetime fitness $\lambda_{ind_{adj}}$ and average annual fitness $\overline{\omega}_{life_{adj}}$ (Fig. 3 B & F, Table 2). In males, we found significant directional selection for earlier relative emergence dates when using lifetime fitness $\lambda_{ind_{adj}}$, or average annual fitness $\overline{\omega}_{life_{adj}}$ (Fig. 3 A, E, Table 2). Because the dataset on adjusted LRS was reduced, we ran the same models (using $\lambda_{ind_{adj}}$ and $\overline{\omega}_{life_{adj}}$) on the reduced datasets to check that the absence of detected selection when using LRS was not due to smaller sample size. Results remained the same in males (directional selection for earlier emergence dates using both $\lambda_{ind_{adj}}$ and $\overline{\omega}_{life_{adj}}$, *Supplementary Material S1*). In females, we found stabilizing selection when considering mean annual fitness $\overline{\omega}_{life_{adj}}$ (P = 0.058), and directional selection for earlier emergence when considering lifetime fitness $\lambda_{ind_{adj}}$ (*Supplementary Material S1*). Finally, our updates of the selection gradients on annual fitness values confirmed directional selection for earlier emergence in females but no selection in males (Fig. 3 G & H, Table 2) as previously reported by Viblanc et al. (2022).

**Strength of selection on emergence date according to fitness metrics**

Standardizing respective values for date of emergence from hibernation to a mean of zero and unit variance (z-scores) allowed comparison of the β selection coefficients obtained for relative emergence date using either annual ($\omega_{an_{adj}}$; updated from Viblanc et al. 2022; using a mixed model with both ID and AGE included as random factors) or lifetime fitness metrics ($\overline{\omega}_{life_{adj}}$, $\lambda_{ind_{adj}}$, LRS$_{adj}$; Fig.4). In females, the standardized selection gradient towards earlier emergence dates was stronger for lifetime than for annual fitness (-0.20 ± 0.07, -0.17 ± 0.07 for average annual fitness $\overline{\omega}_{life_{adj}}$ and lifetime fitness $\lambda_{ind_{adj}}$ respectively *vs*. -0.12 ± 0.04 for annual fitness $\omega_{an_{adj}}$), but was not significant for LRS$_{adj}$ (-0.13 ± 0.08). Despite a higher uncertainty associated with lower sample size in males, similar results were found (-0.36 ± 0.13, -0.33 ± 0.13 for average annual fitness $\overline{\omega}_{life_{adj}}$ and lifetime fitness $\lambda_{ind_{adj}}$ respectively *vs*. -0.15 ± 0.10 for annual fitness $\omega_{an_{adj}}$ and -0.12 ± 0.15 for LRS$_{adj}$).

## DISCUSSION

Whether for male or female Columbian ground squirrels, measures of fitness that included the lifespans of individuals were highly associated with one another (Table 1). Similar results occur in other species, such as Song Sparrows (*Melospiza melodia* Wilson, 1810) and bighorn sheep (*Ovis canadensis* Shaw, 1804) (Reid et al. 2019; Van de Walle et al. 2022). In our study, the associations were still strong, though lower in effect size, when adjustments were made for changes in mean population fitness during individual lifetimes. As elsewhere (Viblanc et al. 2010, 2022; Dobson et al. 2020), we reason that measures adjusted for mean population fitness should more accurately reflect fitness measured on individuals (Fisher 1930). This is because fitness advantages for traits should be compared to other individuals in a population, rather than at other times or places. Associations of the three adjusted lifetime matrix-based $\lambda$ ($\lambda_{ind_{adj}}$), mean lifetime annual fitness ($\bar{\omega}_{life_{adj}}$), and lifetime reproductive success (LRS$_{adj}$) were strong (Table 1). Comparisons of unadjusted and adjusted fitness metrics were non-linear, especially for LRS with the other two fitness measures (Figs. 1 and 2). LRS is a per generation measure of fitness, however, while matrix-based lambda and mean lifetime annual fitness are measured as per year fitness, perhaps contributing to imprecision of comparisons among fitness estimates.

Despite the high correlations of alternative measures of lifetime fitness, different measures produced somewhat different estimates of selection on the date of behavioral emergence from hibernation (Fig. 3, Table 2). Coefficients of selection were strongest for lifetime measures that controlled for mean population fitness ($\lambda_{ind_{adj}}$ and $\bar{\omega}_{life_{adj}}$), for both males and females. The effect sizes of annual fitness ($\omega_{an_{adj}}$) and lifetime reproductive success ($LRS_{adj}$) were much lower, though all metrics had overlapping confidence intervals (Fig. 4). Standardized selection gradients were about 40% greater for the former lifetime estimates than for the annual and LRS measures for females and roughly 100% greater for males. Thus, although the variance in all measures overlapped considerably, there appeared to be substantial differences in effect sizes. In earlier studies (Lane et al. 2012; Viblanc et al. 2022; Thompson et al. 2023), the effect sizes of the selection gradient were estimated using mixed models that controlled for individual identity and age. Effect sizes in those studies were much lower than our estimates of $\lambda_{ind_{adj}}$ and $\bar{\omega}_{life_{adj}}$ and were sometimes not statistically significant, but they were more similar to our estimates of the selection gradient using the annual estimate $\omega_{an_{adj}}$. Thus, it is likely that conclusions about the size or significance of selection gradients will change when using different methods of fitness estimation.

      Naturally, we cannot know which estimates most accurately reflected true fitness. Thus, the choice of a fitness measure best rests on theoretical or practical grounds. An annual fitness measure has the advantage of being convenient for experimental studies where an agent of selection can be experimentally manipulated (Dobson et al. 2020). For species with multiannual lifespans, however, a measure that takes demographic traits into account might be favored (McGraw and Caswell 1996), and can be used to model events that recur during the lifespan. Averaging annual fitness over the lifetime also accounts for repeated events over the lifespan, but does not explicitly take age into account. For long-lived species in the absence of experimental manipulations, either of these lifetime fitness measures ($\lambda_{ind_{adj}}$ and $\bar{\omega}_{life_{adj}}$) should better reflect success in contributing to future generations than any single annual value (Dobson et al. 2020; but see Brommer et al. 2004). The primary advantage of lifetime reproductive success is that the specific ages need not be considered, and reproductive success is simply pooled over the lifespan. For any of the lifetime estimates of fitness that are analyzed for selection gradients, events that recur over the lifespan must also be combined in some way.

Regardless of the lifetime method used, individual fitness is relative to how others fare in the population, so an adjusted measure with respect to mean population fitness might be preferred (Viblanc et al. 2010, 2022).

Regardless of the fitness metric chosen, and whether an adjustment is made for changes in mean population fitness, any study of evolutionary changes needs to focus on heritable traits and the strength of selection on those traits (Endler 1986; Reed et al. 2022). We focused on the date of emergence from hibernation, a significantly heritable trait (males and females, respectively, $h^2$ = 0.34 and 0.22; Lane et al. 2011) that has been shown to be under selection for earlier expression in previous analyses with more limited datasets (Lane et al. 2012; Viblanc et al. 2022; Thompson et al. 2023). Taken together, the results of significant heritability and negative selection gradient suggest that natural selection may favor earlier emergence from spring hibernation, as well as the subsequent and genetically correlated mating season. At the same time, emergence date is a highly phenotypically plastic trait that varies among years and within individuals (Dobson 1988; Lane et al. 2012, 2019; Dobson et al. 2023; Thompson et al. 2023). In addition, males appear to be under sexual selection to emerge above ground well before females (Thompson et al. 2023), perhaps explaining the stronger heritability and selection coefficients for males than females (Lane et al. 2011; Table 2).

Natural selection events that occur only occasionally in a lifetime may not be well tested by annual reproduction and survival. This is because while the event has an immediate influence, individuals that survive will experience other events, such as carry-over effects or life-history tradeoffs, that serve to dilute the strength of selection. For example, those that survive a storm may later have other environmental influences on both survival and reproduction. In this case, the traits that favored the survival of the storm will be passed along (or not passed along) to the next generation partly according to other influences on fitness that come earlier and later in life. In this way, the agents of selection will interact through their influences on the gene pool of the offspring generation. We found that an annual metric produced a lower coefficient of directional selection on the date of emergence from hibernation for males (Fig. 4, Table 2). In this case, extreme years within a lifetime may have contributed unequally to the estimate, while lifetime estimates produced estimates that took several years into account.

In a previous study of Columbian ground squirrels, Dobson et al. (2020) found that correlations among fitness measures were moderately associated. In particular, LRS (unadjusted) seemed a relatively poor measure, since it produced different conclusions about the influence of age at maturity on fitness (Rubach et al. 2020) and exhibited high variability with respect to other fitness measures (Dobson et al. 2020). This is important because of the large number of studies that have used LRS as the primary or sole fitness measure. A recent search of all databases in the Web of Science for "lifetime reproductive success" yielded over 2000 articles (16 October 2024). In both Richardson's ground squirrels (*Urocitellus richardsonii* (Sabine, 1822)) and Song Sparrows, more proximate measures of lifetime fitness were only weakly associated with number of grand-offspring (respectively; Catton and Michener 2016 ; Reid et al. 2019). For Columbian ground squirrels, Viblanc et al. (2022) found variation among annual estimates of selection gradients for fitness regressed on emergence date from hibernation, when the estimates of offspring production were measured at different time points in the reproductive cycle: birth, weaning, or for subadults at a later time. While selection coefficients were similar and generally significant among the different periods of offspring measurement, it was notable that the selection coefficient was not significant for males when offspring were counted as subadults and for females when offspring were counted at weaning. These previous studies and the current one suggest that the choice of a fitness metric is important in studies of natural selection and may markedly influence our conclusions about the strength or direction of selection.

**ACKNOWLEDGMENTS**

We owe great thanks to JO Murie for access to data that he gathered on the Columbian ground squirrel population during 1992-1999. JF Hare, JO Murie, and CS Philson provided excellent constructive comments on versions of the manuscript. We are grateful to Alberta Parks, and Alberta Environment, Fish & Wildlife for granting us access to the study sites and support of the long-term research. The University of Calgary Biogeoscience Institute provided housing at the RB Miller field station during data collection in Sheep River Provincial Park (AB, Canada). We are grateful to EA Johnson and S Vamosi (Directors), J Mappan-Buchanan and A Cunnings (Station Managers) and K Ruckstuhl (Faculty Responsible) for providing us with field camp and laboratory facilities over the years, and to P Neuhaus and K Rucktuhl for their continued support, help, and discussions in the field. We are especially grateful to EA Johnson for his continued support of the ground squirrel long-term research throughout the years. The research was funded by a CNRS Projet International de Coopération Scientifique grant (PICS-07143) and a research grant from the Fondation Fyssen to VAV, by a USA National Science Foundation grant (DEB-0089473) to FSD, by a National Science and Engineering Research Council of Canada grant to DWC (Discovery Grant RGPIN-2018-04354), and by a fellowship grant from the Institute of Advanced Studies of the University of Strasbourg to FSD and VAV. FSD thanks the Région Grand Est and the Eurométropole de Strasbourg for the award of a Gutenberg Excellence Chair during the time of writing. This study is part of the long-term Studies in Ecology and Evolution (SEE-Life) program of the CNRS.

## COMPETING INTERESTS

## DATA AVAILABILITY STATEMENT

Data available on request.

**TABLES**

Table 1. Spearman correlation values (r) between annual and lifetime fitness estimates for male and female Columbian ground squirrels (*Urocitellus columbianus*). (A) Fitness metrics were calculated as lifetime reproductive success (LRS), individual lifetime genotype growth rate ($\lambda_{ind}$; McGraw and Caswell 1996), or mean individual annual fitness $\bar{\omega}_{life}$ (viz. survival + 0.5 x reproduction; Qvarnström et al. 2006). (B) Fitness metrics were adjusted for annual population size or changes in demography over individual lifetimes (see Methods). Sample sizes vary due to the possibility to adjust LRS only for individuals living in cohorts in which all individuals died during the study time and are indicated as N. ***P < 0.001.

|   |   |   | Females | N | Males | N |
|---|---|---|---|---|---|---|
| A | r for fitness metrics (not adjusted for demography) | $\lambda_{ind}, LRS$ | 0.99 *** |  | 0.99 *** |  |
|   |   | $\bar{\omega}_{life}, LRS$ | 0.96 *** | 209 | 0.96 *** | 56 |
|   |   | $\lambda_{ind}, \bar{\omega}_{life}$ | 0.96 *** |  | 0.97 *** |  |
| B | r for fitness metrics (adjusted for demography) | $\lambda_{ind_{adj}}, LRS_{adj}$ | 0.93 *** | 167 | 0.80 *** | 52 |
|   |   | $\bar{\omega}_{life_{adj}}, LRS_{adj}$ | 0.87 *** | 167 | 0.78 *** | 52 |
|   |   | $\lambda_{ind_{adj}}, \bar{\omega}_{life_{adj}}$ | 0.88 *** | 208 | 0.95 *** | 56 |

**Table 2. Directional, stabilizing, or disruptive selection on emergence date from 4 fitness metrics.** Results of linear models (or linear mixed models) testing the effect of linear and quadratic effects of relative emergence dates on lifetime fitness metrics (or annual fitness) in male and female Columbian ground squirrels (*Urocitellus columbianus*) are presented as estimates ± SE, along with the t- and P-values. Bold characters indicate significant terms (at the $P < 0.05$ threshold). Bold italic characters indicate terms that approached significance ($0.05 < P < 0.10$). Results are presented for the full model and for the simple model with the linear effect only. Sample sizes are indicated as number of individuals for the linear models on lifetime fitness metrics and as number of emergence events, number of individuals and number of age classes in the model (as AGE and ID are included as random terms in the linear mixed models). For standardized (to zero mean) selection coefficients, see Figure 4.

|  |  | Directional | Quadratic | n |
|---|---|---|---|---|
| ♂ | $\lambda_{ind_{adj}}$ | **-0.056 ± 0.028, t = -2.02, P = 0.048** | -0.002 ± 0.002, t = -0.91, P = 0.369 | 55 |
|  |  | **-0.034 ± 0.013, t = -2.55, P = 0.014** | Ø |  |
|  | $LRS_{adj}$ | -0.053 ± 0.066, t = -0.80, P = 0.428 | -0.003 ± 0.006, t = -0.49, P = 0.629 | 51 |
|  |  | -0.025 ± 0.032, t = -0.77, P = 0.444 | Ø |  |
|  | $\bar{\omega}_{life_{adj}}$ | *-0.055 ± 0.030, t = -1.81, P = 0.076* | -0.001 ± 0.002, t = -0.54, P = 0.591 | 55 |
|  |  | **-0.041 ± 0.015, t = -2.79, P = 0.007** | Ø |  |
|  | $\omega_{an_{adj}}$ | -0.005 ± 0.010, t = -0.51, P = 0.608 | -0.001 ± 0.001, t = 1.29, P = 0.200 | n = 200, $N_{ID}$ = 80, $N_{age}$ = 9 |
|  |  | -0.012 ± 0.008, t = -1.47, P = 0.143 | Ø |  |
| ♀ | $\lambda_{ind_{adj}}$ | **-0.018 ± 0.007, t = -2.43, P = 0.016** | *-0.002 ± 0.001, t = -2.06, P = 0.040* | 205 |
|  |  | **-0.019 ± 0.007, t = -2.49, P = 0.013** | Ø |  |
|  | $LRS_{adj}$ | -0.030 ± 0.021, t = -1.46, P = 0.148 | -0.004 ± 0.003, t = -1.28, P = 0.201 | 167 |
|  |  | -0.032 ± 0.020, t = -1.59, P = 0.114 | Ø |  |
|  | $\bar{\omega}_{life_{adj}}$ | **-0.022 ± 0.008, t = -2.85, P = 0.005** | **-0.003 ± 0.001, t = -2.49, P = 0.013** | 206 |
|  |  | **-0.022 ± 0.008, t = -2.90, P = 0.004** | Ø |  |
|  | $\omega_{an_{adj}}$ | **-0.014 ± 0.004, t = -3.18, P = 0.002** | *-0.001 ± 0.000, t = -1.72, P = 0.087* | n = 805, $N_{ID}$ = 246, $N_{age}$ = 14 |
|  |  | **-0.012 ± 0.004, t = -2.87, P = 0.004** | Ø |  |

**Figure captions.**

Figure 1. Comparisons of lifetime fitness metrics for Columbian ground squirrels (*Urocitellus columbianus*), not adjusted for mean fitness during each individual's lifespan. LRS is the lifetime sum of offspring that survive to one year from conception. $\lambda_{ind}$ is the dominant eigenvalue of the pre-breeding Leslie matrix. $\bar{\omega}_{life}$ is the adjusted average annual fitness over an individual's life. Metrics for females appear to the left and those for males on the right.

Figure 2. Comparisons of lifetime fitness metrics for Columbian ground squirrels (*Urocitellus columbianus*), adjusted for mean fitness during each individual's lifespan. LRS$_{adj}$ is the adjusted lifetime sum of offspring that survive to one year from conception. $\lambda_{ind_{adj}}$ is the adjusted dominant eigenvalue of the pre-breeding Leslie matrix. $\bar{\omega}_{life_{adj}}$ is the average adjusted annual fitness over an individual's life. Metrics for females appear to the left and those for males on the right.

Figure 3. Lifetime fitness metrics for Columbian ground squirrels (*Urocitellus columbianus*, adjusted for mean fitness, regressed on the annual mean centered date of emergence above ground from hibernation. $\lambda_{ind_{adj}}$ is the adjusted dominant eigenvalue of the pre-breeding Leslie matrix. LRS$_{adj}$ is the adjusted lifetime sum of offspring that survive to one year from conception. $\bar{\omega}_{life_{adj}}$ is the average adjusted annual fitness over an individual's life. $\omega_{an_{adj}}$ is annual fitness divided by the mean annual fitness for a given year. Metrics for males appear to the left and those for females on the right.

Figure 4. Standardized selection coefficients (to zero mean) for the regression of fitness metrics on the annual mean centered data of emergence above ground from hibernation by Columbian ground squirrels (*Urocitellus columbianus*). $\omega_{an_{adj}}$ is annual fitness divided by the mean annual fitness for a given year. $\bar{\omega}_{life_{adj}}$ is the average adjusted annual fitness over an individual's life. $\lambda_{ind_{adj}}$ is the adjusted dominant eigenvalue of the pre-breeding Leslie matrix. LRS$_{adj}$ is the adjusted lifetime sum of offspring that survive to one year from conception. Metrics for males appear the top and those for females lower down. Whiskers indicate 95% confidence intervals about the mean.

# FIGURES

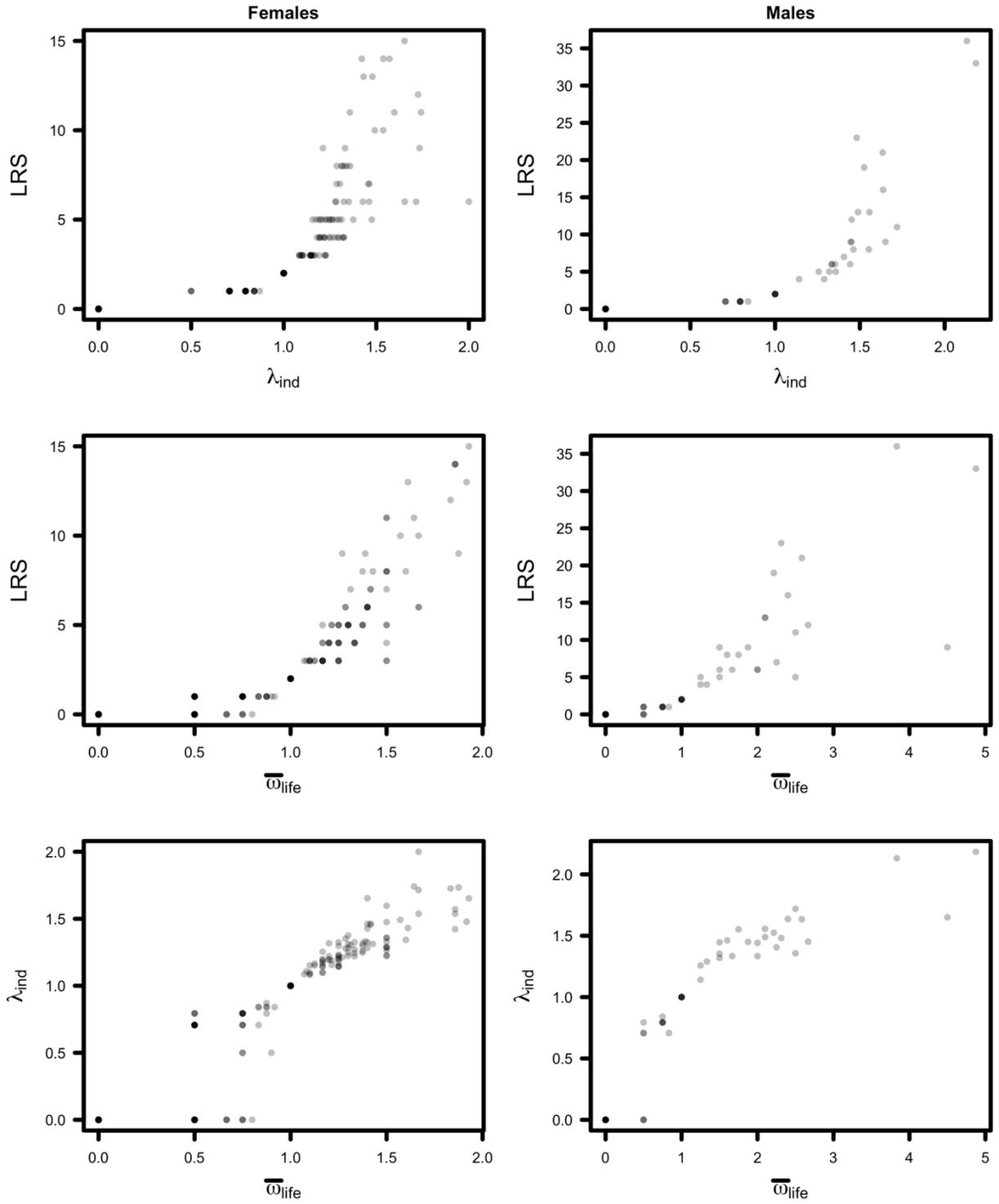

Figure 1.

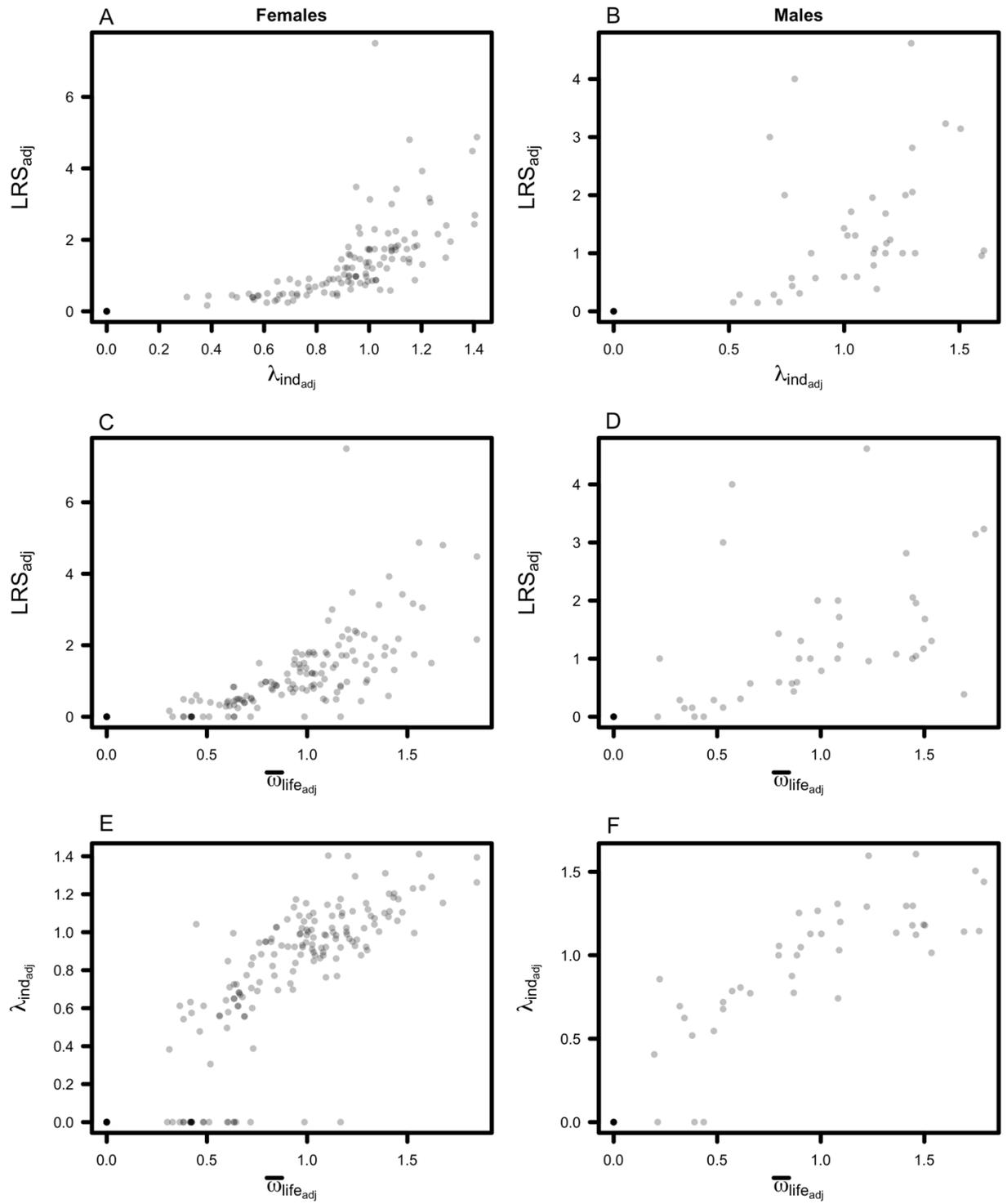

Figure 2.

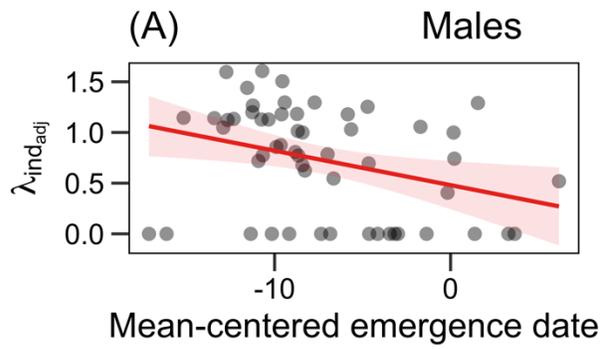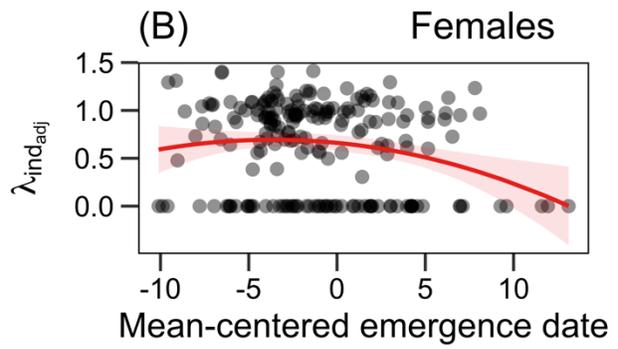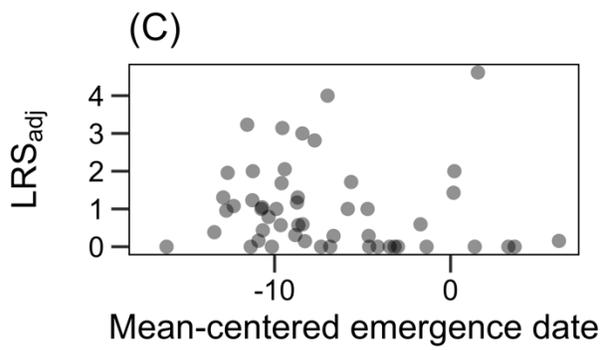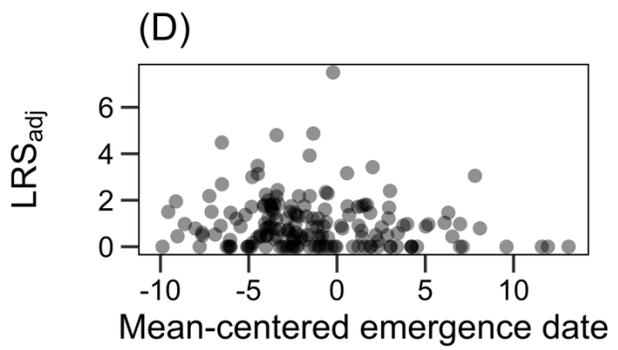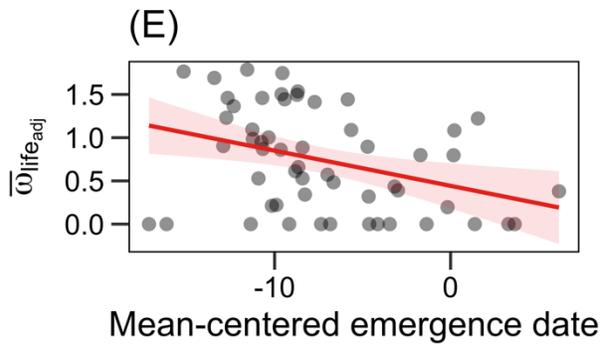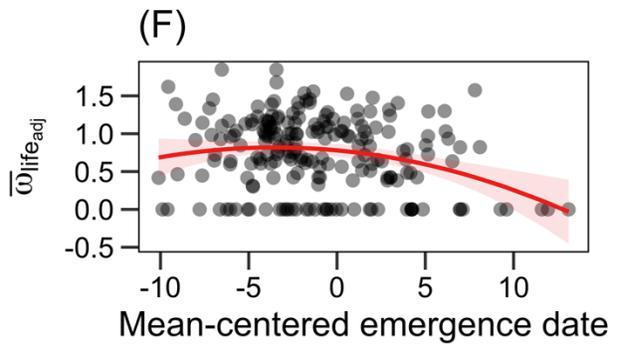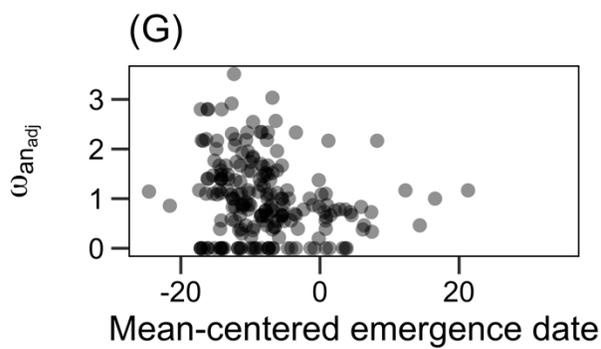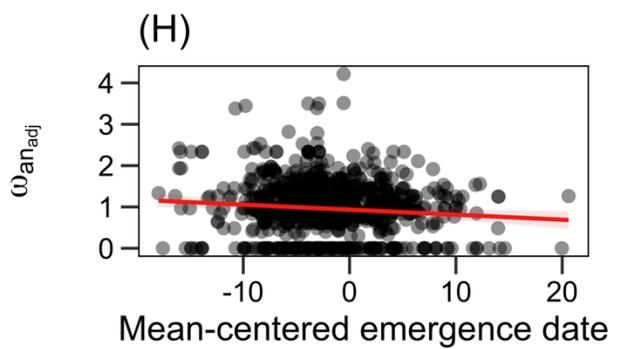

Figure 3

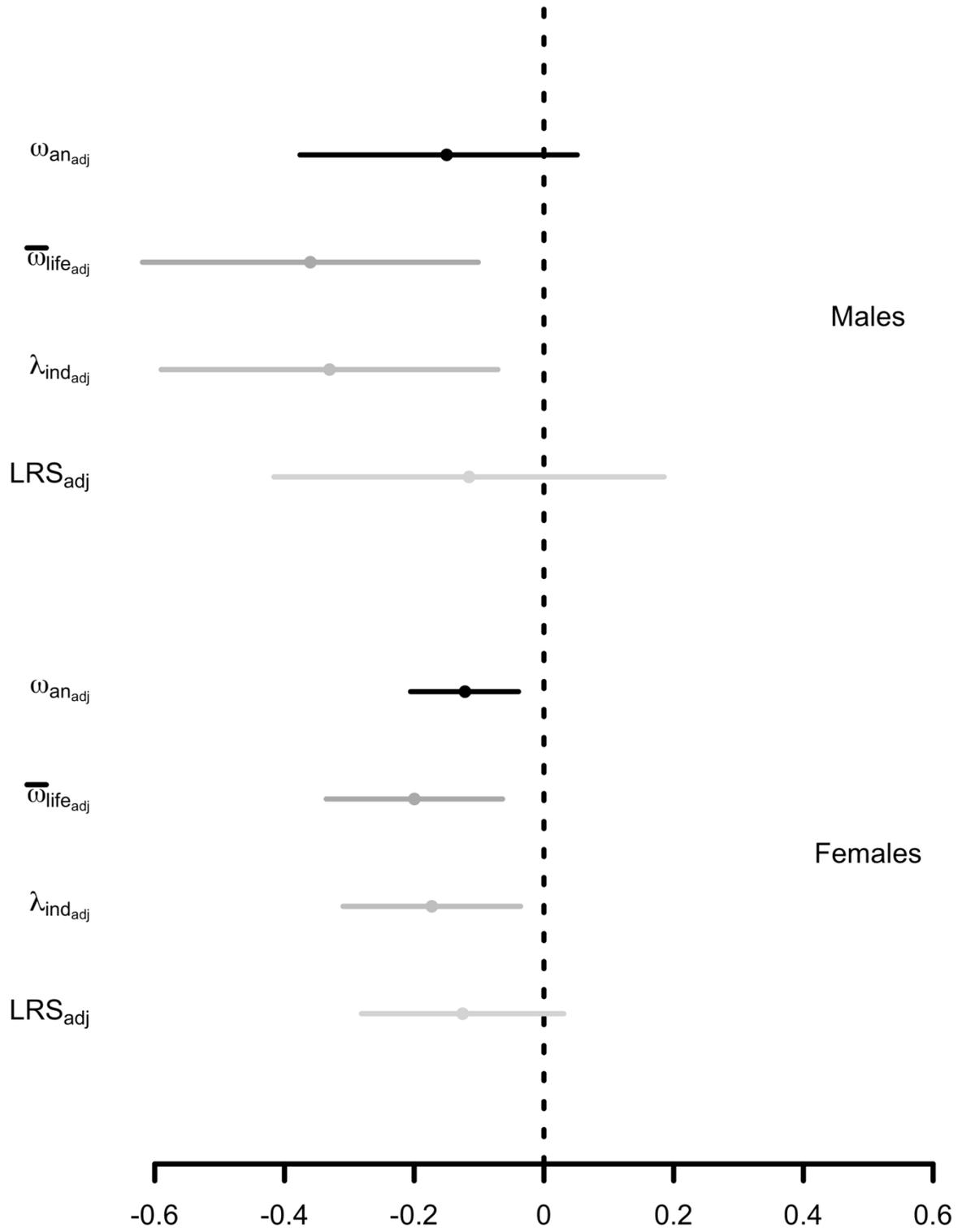

Figure 4.